# The coverage of Microsoft Academic: Analyzing the publication output of a university

Sven E. Hug[1,2,*,†] and Martin P. Brändle[3,4,†]

[1] Social Psychology and Research on Higher Education, ETH Zurich, D-GESS, Muehlegasse 21, 8001 Zurich, Switzerland
[2] Evaluation Office, University of Zurich, 8001 Zurich, Switzerland
[3] Zentrale Informatik, University of Zurich, 8006 Zurich, Switzerland
[4] Main Library, University of Zurich, 8057 Zürich, Switzerland
[†] Both authors contributed equally.
[*] Corresponding author. Tel.: +41 44 632 46 85, Fax: +41 44 634 43 79, Email: sven.hug@gess.ethz.ch

**Abstract** This is the first detailed study on the coverage of Microsoft Academic (MA). Based on the complete and verified publication list of a university, the coverage of MA was assessed and compared with two benchmark databases, Scopus and Web of Science (WoS), on the level of individual publications. Citation counts were analyzed, and issues related to data retrieval and data quality were examined. A Perl script was written to retrieve metadata from MA based on publication titles. The script is freely available on GitHub. We find that MA covers journal articles, working papers, and conference items to a substantial extent and indexes more document types than the benchmark databases (e.g., working papers, dissertations). MA clearly surpasses Scopus and WoS in covering book-related document types and conference items but falls slightly behind Scopus in journal articles. The coverage of MA is favorable for evaluative bibliometrics in most research fields, including economics/business, computer/information sciences, and mathematics. However, MA shows biases similar to Scopus and WoS with regard to the coverage of the humanities, non-English publications, and open-access publications. Rank correlations of citation counts are high between MA and the benchmark databases. We find that the publication year is correct for 89.5% of all publications and the number of authors is correct for 95.1% of the journal articles. Given the fast and ongoing development of MA, we conclude that MA is on the verge of becoming a bibliometric superpower. However, comprehensive studies on the quality of MA metadata are still lacking.

# Introduction

The Microsoft Academic Graph (MAG) was established in June 2015 (Microsoft, 2017a) and models "the real-life academic communication activities as a heterogeneous graph" (Sinha et al., 2015, p. 244). It gets most of its data from web pages indexed by Bing (Sinha et al., 2015) and is updated on a weekly basis (Microsoft, 2017a). MAG data can be accessed via the Microsoft Academic search engine[1] or via the Academic Knowledge API (AK API)[2]. We refer to these services and to the constituents of these services as Microsoft Academic (MA), which was available as a preview beginning in February 2016 (Microsoft, 2017b) and was officially launched as version 2.0 in July 2017 (Microsoft, 2017c). MA is evolving quickly and several new features have been implemented since our first examination of the database (Hug, Ochsner, & Brändle, 2017). For example, a social network for academics and a search function for graph patterns have been integrated. Furthermore, aggregated citation counts for authors, institutions, fields, journals, and conferences have been added to the API. Most interestingly, citation contexts for references are now retrievable, which allows for the calculation of advanced indicators as suggested by Waltman (2016) and paves the way for the study of citation acts (Bertin, 2008; Bertin, Atanassova, Sugimoto, & Lariviere, 2016). MA is also progressing quickly in terms of coverage. According to the development team of MA, the database expanded from 83 million records in 2015 (Sinha et al., 2015) to 140 million in 2016 (Wade, Wang, Sun, & Gulli, 2016) and 168 million in early 2017 (A. Chen, personal communication, March 31, 2017). It is currently growing by 1.3 million records per month (Microsoft Academic, 2017).

The predecessor of MA, Microsoft Academic Search, was decommissioned towards the end of 2016 and attracted little bibliometric research. Harzing (2016) identified only six journal articles related to Microsoft Academic Search and bibliometrics. In contrast, MA has already spurred great interest in a short period of time and triggered several studies that focus on bibliometric topics, such as four studies on visualization and mapping (De Domenico, Omodei, & Arenas, 2016; Portenoy, Hullman, & West, 2016; Portenoy, & West, 2017, Tan et al., 2016). Furthermore, there are eleven studies that deal with the development of indicators and algorithms (Effendy & Yap, 2016; Effendy & Yap, 2017; Herrmannova & Knoth, 2016b; Luo, Gong, Hu, Duan, & Ma, 2016; Medo & Cimini, 2016; Ribas, Ueda, Santos, Ribeiro-Neto, & Ziviani, 2016; Sandulescu & Chiru, 2016; Wesley-Smith, Bergstrom, & West, 2016;

---

[1] https://academic.microsoft.com
[2] https://www.aka.ms/AcademicAPI

Vaccario, Medo, Wider, & Mariani, 2017; Wilson, Mohan, Arif, Chaudhury, & Lall, 2016; Xiao et al., 2016). Finally, there are four studies that assess the potential of MA for evaluative bibliometrics. Hug et al. (2017) examined the strengths and weaknesses of the AK API from the perspective of bibliometrics and calculated normalized indicators (i.e. average-based and distribution-based indicators). Harzing (2016) and Harzing and Alakangas (2017a) compared publication and citation coverage of MA with Scopus, Web of Science (WoS), and Google Scholar. Herrmannova and Knoth (2016a) compared features of the metadata stored in Mendeley, COnnecting REpositories (CORE), and the MAG. They also compared rankings of universities and journals based on MAG data with the SCImago Journal & Country Rank and the Webometrics Ranking of World Universities.

In evaluative bibliometrics, it is crucial to know how well a given database covers publications in order to decide whether it is valid for citation analysis (Mongeon & Paul-Hus, 2016). In the studies of Harzing (2016), Harzing and Alakangas (2017a), and Herrmannova and Knoth (2016a), the publication coverage of MA was addressed, but the results were inconclusive for two reasons. First, these studies provided little empirical evidence, since one study analyzed a very small sample size (Harzing, 2016) and the two large-scale studies did not investigate publication coverage in detail (Harzing & Alakangas, 2017a; Herrmannova & Knoth, 2016a). Harzing and Alakangas (2017a) calculated the average number of papers of 145 academics in MA, Scopus, WoS, and Google Scholar. They found that, on average, MA reports more papers per academic (137) than Scopus (96) and WoS (96) and less than Google Scholar (155). They provided no further information on publication coverage. From a methodological standpoint, the drawback of Harzing and Alakangas' (2017a) study is that publications were collected by author queries and not on the level of individual and verified publications, as required by Moed (2005). Herrmannova and Knoth (2016a) analyzed the number of DOIs stored in MA, Mendeley, and CORE and found that there are 35.5 million unique DOIs in MA. They also analyzed the number of publications assigned to the different fields in MA. However, according to Hug et al. (2017), raw field information from MA cannot readily be used for bibliometric purposes. Second, in the studies Harzing and Alakangas (2017a) and Herrmannova and Knoth (2016a), the coverage was analyzed in relation to other databases only. Hence, these studies did not assess how well actual publication lists of scholars, institutes, or universities are represented in MA. Put differently, there are no studies on the recall of an actual publication list, where recall is defined as the fraction of relevant documents that are retrieved (Manning, Raghavan, & Schütze, 2008).

The main goal of the present study is to assess the publication coverage of MA in relation to an actual publication list. We will analyze the coverage of a verified publication list of a university in MA and two benchmark databases (Scopus and WoS) on the level of individual publications. In addition, we will analyze citation counts and examine issues related to data retrieval and data quality. The following research questions will be addressed:

- What is the coverage of the publication list in MA, Scopus, and WoS with regard to document type, publication language, access status (open or restricted), publication year, and research field?
- How do citations correlate between MA, Scopus, and WoS?
- What are the citations per publication and what is the share of uncited publications in the three databases?
- What is the quality of the metadata in MA with respect to DOI coverage, number of authors per paper, and publication year?

The remainder of the article is organized as follows. The Method section is organized in two parts. In the first part, the Zurich Open Archive and Repository (ZORA) is described, from which the publications for this study are drawn. This includes information about ZORA in general, the definition of the publication sets, specifications of the repository software, and a description of data harvesting from the benchmark databases. In the second part, a Perl script is specified, which retrieves metadata of ZORA items from MA in two modes (title_exact and title_word) and evaluates the matching quality of the retrieved data. In the Results sections, the performance of the Perl script with respect to retrieval and matching quality, the evaluation of the MA metadata, the assessment of the coverage, and the analysis of citation counts are presented. Finally, the results are discussed and a conclusion is provided.

## Method

### Zurich Open Archive and Repository (ZORA)

The publications for the present study were drawn from ZORA[3], an open archive and repository in which the University of Zurich (UZH) documents its publication output. Research at UZH covers all broad disciplinary areas, although the number of publications in engineering is lower than in other disciplines. ZORA was established in 2006, and since 2008, the UZH has required its researchers to deposit metadata of their publications in ZORA, including full text whenever possible. According to the ZORA regulations (Main Library of the University of Zurich, 2017), researchers feed their publications into the repository and "all publications are checked and completed by the ZORA editorial team in accordance with the faculties. The focus of this work is on the quality of bibliographic data […] and on the copyright situation." For example, the publication year of a submitted record is compared with the original publication by the editorial team, and items published "online first" are checked automatically in Crossref daily. Therefore, the repository offers a complete and verified publication list, which, according to Moed (2005), is a crucial first step for evaluative bibliometrics. In addition to the usual bibliographic data, the metadata in ZORA also includes further information such as publication language, access status (open or restricted), affiliation to UZH institutes, and document type. These four variables plus the publication year were used to analyze the coverage of ZORA items in MA, Scopus, and WoS.

---
[3] https://www.zora.uzh.ch

When data for the present study were collected in October 2016, ZORA contained a total of 91,215 items. We refer to this publication set as $ZORA^{total}$. This set was used to assess the retrieval and matching quality of the Perl script and to evaluate the metadata of MA. To compare MA with Scopus and WoS from the perspective of evaluative bibliometrics, a subset of $ZORA^{total}$ was used, which is referred to as $ZORA^{2008-2015}$. The subset comprised 62,791 items and differed in three ways from $ZORA^{total}$. First, it encompassed the publication years 2008 to 2015, because data entry in ZORA is mandatory only for publications since 2008 and data collection and verification of the publication year 2016 will only be finished in 2017. Second, only publications from researchers at institutes were included, since researchers at other organizational units of UZH (e.g., competence centers, research priority programs) are not required to feed their publications into the repository. Third, only main document types used in scholarly communication and evaluative bibliometrics (journal articles, conference items, monographs, book sections, edited volumes) were included because the acceptance of further document types, such as dissertations and scientific reports, is an open question in evaluative citation analysis (Prins, Costas, van Leeuwen, & Wouters, 2016).

ZORA is based on the open-source EPrints repository software (version 3.3)[4]. For relevance-ranked searches, the UZH incarnation runs the Xapian search engine as an add-on. Plug-ins and scripts extend the standard EPrints functionality of ZORA. Scopus citation counts are harvested using the *citation count dataset and import plug-in* developed by the Queensland

---

[4] http://www.eprints.org/

University of Technology[5] and modified by UZH. This plug-in queries the Scopus API[6] sequentially—first with Scopus EID and second with the DOI for any document type, followed by PubMed ID, ISBN, and bibliographic metadata for journal articles, book sections, conference items, monographs, and edited volumes. Daily batches of ZORA item data (about 3,000 items) are sent for all items to be processed once a month. WoS citation counts are harvested using a script developed at the UZH that accesses WoS via the Link Article Match Retrieval API[7]. This script sends queries for the five document types mentioned above by using WoS UT number, DOI, PubMed ID, ISBN, and bibliographic metadata. All ZORA items are processed once a week.

**Perl script for retrieving and matching Microsoft Academic data**

The AK API offers four REST endpoints to access MAG data.[8] For this study, the *Evaluate* endpoint was used, which retrieves metadata of publications (called *entity attributes* in MA) from the MAG based on a query expression. An *Evaluate* request returns one or several items that match the query expression. We limited the maximum number of items that could be returned per query to ten, as our test runs had shown that this was sufficient for the purpose of the study. The AK API assigns to each returned item a natural log probability value to indicate the quality of the match and, by default, ranks the returned items by descending probability in the result set. To build query expressions, we used information from the titles of ZORA items since DOIs cannot be used in query expressions (Hug et al., 2017). A Perl script (academic_search) was written which relies on the *Evaluate* method to query the MAG with data from an EPrints repository, saves the JSON results, evaluates whether the results sent by the AK API match the ZORA item (see below), and creates reports in XML and CSV formats. The script allows for four different retrieval modes, of which the following two are relevant for this study: title_words and title_exact.

The title_words (ti_wo) mode takes a stop word-filtered list of title words from the ZORA Xapian index and constructs an AND-nested query expression using the MA entity attribute "W" (W = words from paper title/abstract for full text search). For example, the query expression of the record with the title "HEE-GER: a systematic review of German economic evaluations of health care published 1990-2004" is constructed as:

---

[5] https://github.com/QUTlib/citation-import
[6] http://api.elsevier.com/content/search/scopus
[7] http://ipscience-help.thomsonreuters.com/LAMRService/WebServicesOverviewGroup/overview.html
[8] For an overview of the AK API see https://docs.microsoft.com/en-us/azure/cognitive-services/academic-knowledge/home

And(And(And(And(And(And(And(And(And(W='care',W='economic'),W='evaluations'),W=' ger'),W='german'),W='health'),W='hee'),W='published'),W='review'),W='systematic'). The ti_wo mode applies a list of about 1,500 stop words in English, German, French, Italian, and Spanish, including stop words that were supplied by the development team of MA (D. Eide, personal communication, October 10, 2016). It further filters out numbers. Since the query expression uses the apostrophe as delimiter, the ti_wo mode takes as query word only the part before or after the apostrophe if the index term from Xapian contains one. It takes the longer part of the string.

The title_exact (ti_ex) mode creates an exact title query from the publication title stored in ZORA by transforming the title to lowercase, filtering out special characters, and removing superfluous whitespace. The query expression uses the MA entity attribute "Ti" (Ti = paper title) and, as shown with the example above, is constructed as: Ti='hee ger a systematic review of german economic evaluations of health care published 1990 2004'. The performance of the two retrieval modes were assessed from the perspective of recall, precision, and the $F_1$ score. Recall (R) is defined as the fraction of relevant documents that are retrieved, precision (P) as the fraction of retrieved documents that are relevant, and the $F_1$ score as the harmonic mean of precision and recall (Manning et al., 2008).

The Perl script allows for processing the whole publication set of an EPrints repository or specifying individual publications by their EPrints IDs. Furthermore, there is a restart option that renders it possible to continue querying the AK API at the point where the script was interrupted. Mapping of institutes to research fields is performed by a CSV file read by the script. The query parameters can be configured in a configuration file. The following parameters were used: count=10 (the maximum number of items returned by the AK API per query), model=latest, offset=0. The returned items were ranked by descending probability (default setting in the AK API). The following MA entity attributes were retrieved: Id, Ti, Y, D, CC, ECC, AA.AuN, AA.AuId, AA.AfN, AA.AfId, F.FN, F.FId, J.JN, J.JId, C.CN, C.CId, RId, and the extended metadata attributes E, which (among other things) contains the DOI. After receiving the data from the AK API, the algorithm evaluates each returned item by comparing MA entity attributes with ZORA metadata. A returned item is considered a match if at least one of the following three match types apply: *doi* – DOI in MA and ZORA are equal; *title* – the cleaned title string from the ti_ex mode is identical to the MA title; *bib* – bibliographic data (journal title, volume, issue, first page) is identical. The algorithm then selects the returned item with the most reliable match based on the following priority to determine the matched items: *doi* is considered to be most reliable to determine a matched item due to the uniqueness of the DOI, *title* is considered to be less reliable, and *bib* the least reliable. After selecting the most reliable matches, the parsed MA data and ZORA metadata are stored together with the following two variables: the rank of the matched item in the AK API result set (1-10) and the match type of the matched item (*doi*, *title*, *bib*). These variables are employed to assess the matching quality of ZORA and MA data as well as to assess the performance of the retrieval modes. The Perl script and its associated files are available on GitHub.[9]

## Results

### Performance of the retrieval modes

*Recall, precision, and $F_1$ score*

While 49.7% of the items from ZORA$^{total}$ were retrieved from MA both with the ti_ex and ti_wo modes (45,378 items), 1.4% could be retrieved with the ti_ex mode only (1,319 items) and 1.7% with the ti_wo mode only (1,534 items). Hence, the recall of the ti_ex mode was slightly lower (51.2%) than that of the ti_wo mode (51.4%). The combination of the results of the two modes yielded an overall recall of 52.9% (48,231 items). The precision was calculated by dividing the sum of matched items by the sum of returned items (see Table 1). Since a maximum of ten items was returned per query, the precision scores represent upper estimates. The precision of the ti_ex mode (0.897) was considerably higher than the precision of the ti_wo mode (0.703). For 922 items retrieved by both the ti_ex and ti_wo modes, multiple MA IDs were obtained (see Table 2). These items most likely represented false positives and were thus subtracted from the matched items. Based on the corrected matched items, a corrected precision was calculated, which was 0.879 for ti_ex and 0.689 for ti_wo (see Table 1). Based on recall and the corrected precision, the $F_1$ score was calculated, which yielded a score of 0.647 for the ti_ex mode and 0.594 for the ti_wo mode.

---

[9] https://github.com/eprintsug/microsoft-academic

Table 1  Recall and precision of the retrieval modes title_exact and title_word based on ZORA[total]

| Retrieval mode | Matched items | Corrected matched items | Returned items | R | P | P corrected | F$_1$ corrected |
|---|---|---|---|---|---|---|---|
| ti_ex | 46,697 | 45,775 | 52,067 | 0.512 | 0.897 | 0.879 | 0.647 |
| ti_wo | 46,912 | 45,990 | 66,771 | 0.514 | 0.703 | 0.689 | 0.594 |
| Combined | 48,231 | 47,309 | 59,419[a] | 0.529 | 0.812 | 0.796 | 0.641 |

*Note:* R = recall (in % of ZORA[total]). P = precision (in % of returned items). ti_ex = retrieval mode title_exact. ti_wo = retrieval mode title_word. Values for P and P corrected represent upper estimates since a maximum of 10 items could be returned per query. a = average of returned items by ti_ex and ti_wo.

*Rank in the AK API result set*

Almost all matched items ranked among the top three in the result set of the AK API (ti_ex: 99.4% of the matched items; ti_wo: 98.4%). The ti_ex mode was slightly more precise than the ti_wo mode, as the share of matched items on the first place in the result set of the AK API revealed (95.8% vs. 92.5%).

**Performance of the matching algorithm**

Of the matched items (*n* = 48,231), 69.9% were matched by *doi*, 29.7% by *title*, and 0.4% by *bib*. To assess the precision of the matching algorithm in the Perl script, the differences in the results, which were both retrieved by the ti_ex and the ti_wo modes (*n* = 45,378), were used. We assume that if a ZORA item got the same MA ID by the ti_ex and ti_wo modes, the matching algorithm selected the correct item. As shown in Table 2, this was the case for 98% of the matched items. However, 922 of the matched items (2%) had different MA IDs and, hence, represented most likely false positives. Of the 511 items that were selected by the same match type but returned different MA IDs (see Table 2), 280 items were matched by *doi* and 231 were matched by *title*. While it is probable that there are publications that have the exact same title, the items matched by *doi* seem to indicate duplicate records in MA. Inspection of some of the 280 items matched by *doi* indeed revealed that the publications were the same but with slight variations in the MA title. Given that 29,586 items were matched by *doi* in both retrieval modes, the share of false positive DOI matches in our analysis was 0.9% (i.e. 280 of 29,586 items).

Table 2  Differences and commonalities between the retrieval modes title_exact and title_word regarding MA ID and match type based on ZORA$^{total}$

|  | MA ID | | | |
| --- | --- | --- | --- | --- |
| Match type | Same in ti_ex and ti_wo | | Different in ti_ex and ti_wo | |
|  | No. | % | No. | % |
| Same in ti_ex and ti_wo | 40,588 | 89.4 | 511 | 1.1 |
| Different in ti_ex and ti_wo | 3,898 | 8.6 | 411 | 0.9 |
| Total | 44,456 | 98.0 | 922 | 2.0 |

*Note:* ti_ex = retrieval mode title_exact. ti_wo = retrieval mode title_word. % = in % of items retrieved by both the modes ti_ex and ti_wo ($n$ = 45,378 items).

**Quality of Microsoft Academic metadata**

*Publication year*

Comparing the publication years of the matched items ($n$ = 48,231) between MA and ZORA revealed that 89.5% of the items had identical publication years, 7.0% differed by ±1 year (+1 year in MA: 2.1%; -1 year: 4.9%), and 3.5% featured larger differences (> +1 year: 1.7%; < -1 year: 1.8%). The publication years of the matched items ranged from 1966 to 2017 in ZORA$^{total}$ and from 1866 to 2017 in MA.

*Number of authors*

Comparing the number of authors of the matched journal articles ($n$ = 42,201) between MA and ZORA showed that 95.1% of the articles had identical author numbers, 1.7% differed by ±1 author (+1 author in MA: 1.0%; -1 author: 0.7%), and 3.2% featured larger differences (> +1 author: 2.4%; < -1 author: 0.8%). The number of authors of further publication types could not be analyzed, as ZORA only provides reliable author counts for journal articles.

*DOI*

In ZORA$^{total}$, 51.6% of the items were equipped with a DOI, and a DOI was available for 77.6% of the matched items in MA (see Table 3). In the natural sciences, engineering/technology, medical/health sciences, and agricultural sciences, the proportion of matched items equipped with a DOI in MA was high (76.7% to 91.4%). This proportion was considerably lower in the social sciences (60.5%) and the humanities (37.5%). Only 12.6% of the matched items that had a DOI in ZORA$^{total}$ did not have one in MA. All DOIs of the matched items were valid in MA (i.e., they started with the prefix "10").

Table 3   Availability of DOIs in ZORA$^{total}$ and MA by research field

|  | ZORA$^{total}$ | | MA | |
|---|---|---|---|---|
|  | *n* | % DOI | *N* | % DOI |
| Total | 91,215 | 51.6 | 48,231 | 77.6 |
| Natural Sciences | 15,270 | 72.9 | 11,274 | 76.7 |
| Engineering & Technology | 1,071 | 96.4 | 1,008 | 91.4 |
| Medical & Health Sciences | 32,893 | 75.5 | 25,349 | 81.0 |
| Agricultural Sciences | 5,237 | 63.1 | 3,175 | 81.3 |
| Social Sciences | 19,809 | 26.7 | 6,776 | 60.5 |
| Humanities | 15,854 | 7.9 | 1,468 | 37.5 |
| Other | 5,209 | 49.4 | 3,543 | 71.8 |

*Note*: A publication can be assigned to multiple fields. % DOI = percentage of items equipped with a DOI. Other = publications not assigned to a field since they belong to special collections.

**Coverage**

*Overall and unique coverage of ZORA$^{2008-2015}$*

Scopus covers 57.9% of the ZORA$^{2008-2015}$ publications, MA 56.6%, and WoS 52.6%. There are 2,781 items from ZORA$^{2008-2015}$ that are exclusively covered by MA, 1,655 by Scopus, and 508 by WoS (see Table 4). To examine the coverage of the databases without document type restrictions, ZORA$^{2008-2015}$ was expanded beyond main document types, and 7,164 items registered as further document types were added (see Fig. 2). This expanded publication set consists of 69,955 items. MA covers 1,177 items of the further document types (see Fig. 2), while the benchmark databases cover none, as their indexing policy does not include the document types listed in Fig. 2 (Clarivate, 2017; Elsevier, 2017). As a result, MA covers 52.5% (36,734 items) of the expanded publication set, Scopus 52.0% (36,351 items) and WoS 47.2% (33,000 items). These results show that, overall, MA and Scopus cover ZORA to a similar extent. Of the three databases, MA covers the most items exclusively. MA and Scopus both outperform WoS in terms of overall as well as unique coverage.

Table 4  Overall and unique coverage of ZORA$^{2008\text{-}2015}$

|  | MA | | Scopus | | WoS | |
|---|---|---|---|---|---|---|
|  | No. | % | No. | % | No. | % |
| Overall coverage | 35,557 | 56.6 | 36,351 | 57.9 | 33,000 | 52.6 |
| Unique coverage | 2,781 | 4.4 | 1,655 | 2.6 | 508 | 0.8 |

*Note:* % = percentage of items from ZORA$^{2008\text{-}2015}$ covered by database ($N$ = 62,791). Unique coverage = items from ZORA$^{2008\text{-}2015}$ covered exclusively by one database.

*Main document types*

Scopus covers journal articles from ZORA$^{2008\text{-}2015}$ slightly better than MA and WoS (see Fig. 1). However, MA clearly surpasses Scopus and WoS with respect to book-related document types (i.e., monographs, edited volumes, book sections). For example, MA covers 1.5 times more book sections and 6 times more edited volumes than Scopus. And MA covers 2.2 times more book sections and 4.6 times more monographs than WoS. Despite that, the coverage of book-related items is still very low in MA and reaches a maximum of 15.6% (edited volumes). In contrast, MA covers conference items well (47.8%). Compared to MA, the coverage of conference items is somewhat lower in Scopus (38.5%) and considerably lower in WoS (20.6%).

**Fig. 1**   Coverage of ZORA$^{2008\text{-}2015}$ by main document types

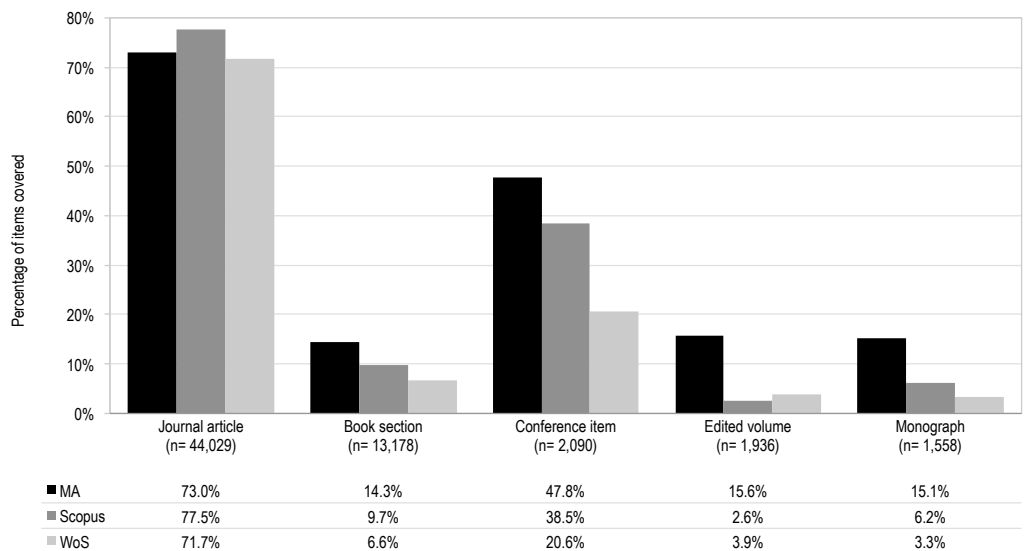

*Further document types*

In addition to the main document types, MA covers further document types (see Fig. 2) which are not indexed by the benchmark databases. In particular, MA covers working papers to a substantial extent. None of the 1,260 newspaper articles in the publication set can be found in MA. The other four document types have a very low representation in MA (ranging from 2.7% for habilitations to 12.9% for dissertations).

**Fig. 2**     Coverage of further document types

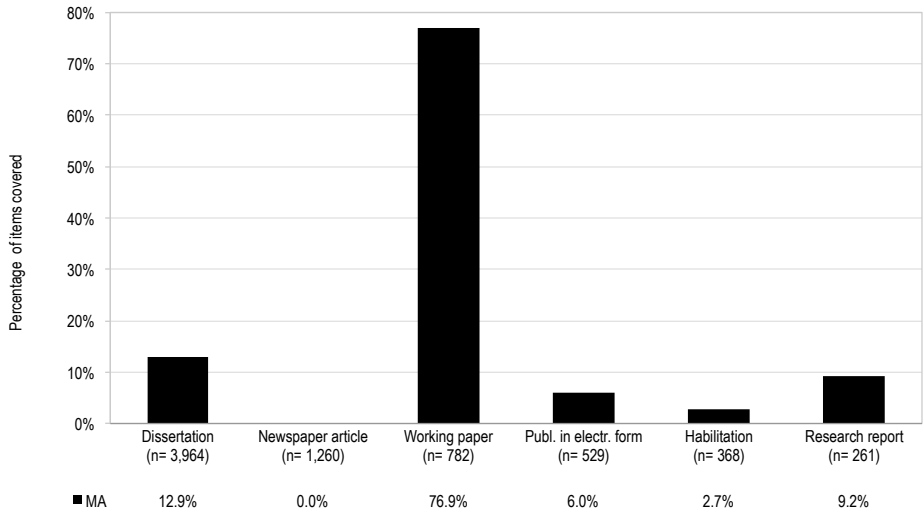

*Note:* The publication set used in Fig. 2 ($N = 7,164$) was derived from ZORA$^{total}$ by the same criteria as ZORA$^{2008-2015}$, but instead of selecting main document types, document types specific to ZORA and/or belonging to grey literature were selected. In total, MA covers 1,177 items of the publication set used in Fig. 2.

*Publication language and access status*

In ZORA$^{2008-2015}$, 61% of the publications are in English, 33% are in German, French, or Italian (official languages of Switzerland), 1% are in 49 other languages, and for 5% of the publications language information is missing. We refer to items in languages other than English as *non-English*. All three databases feature a high coverage of English publications and a low coverage of non-English publications (see Table 5). This is due to a high coverage of journal articles and conference items (see Fig. 1), which are mainly written in English, and a low coverage of book-related items, which are mostly published in the official languages of Switzerland. Regarding the access status of full texts, all three databases have a high coverage of items that are not publicly accessible and, comparatively, a lower coverage of open-access publications (see Table 5). Items without full texts are covered the least, as many of these items are book-related.

Table 5   Coverage of ZORA$^{2008-2015}$ by publication language and access status

|  | ZORA$^{2008-2015}$ n | MA % | Scopus % | WoS % |
|---|---|---|---|---|
| Publication language |  |  |  |  |
| English | 38,551 | 82.8 | 83.2 | 78.2 |
| Non-English | 20,855 | 8.2 | 11.2 | 5.1 |
| Missing | 3,385 | 56.4 | 57.2 | 52.4 |
| Access status of text |  |  |  |  |
| Public | 20,139 | 65.1 | 66.1 | 59.8 |
| Not public | 16,434 | 75.8 | 80.3 | 75.2 |
| No text deposited | 26,218 | 38.1 | 37.6 | 32.8 |

*Note:* Not public = full text only available to members of UZH due to copyright or embargo restrictions. No text deposited = items in ZORA for which no full text has been deposited.

*Publication year*

From 2008 to 2015, the three databases show a similar linear increase in the coverage of ZORA$^{2008-2015}$ (see Fig. 3). Hence, there seems to be no data gap in MA with respect to the publication years 2008 to 2015. The data in Fig. 3 could mistakenly be interpreted as evidence for a substantial expansion of the coverage of the three databases. Instead, researchers at UZH published more journal articles in 2015 (67.3% of all items in that year) than in 2008 (57.9% of all items in that year), which have a better coverage than other document types (see Fig. 1).

**Fig. 3**   Coverage of ZORA$^{2008-2015}$ by publication year

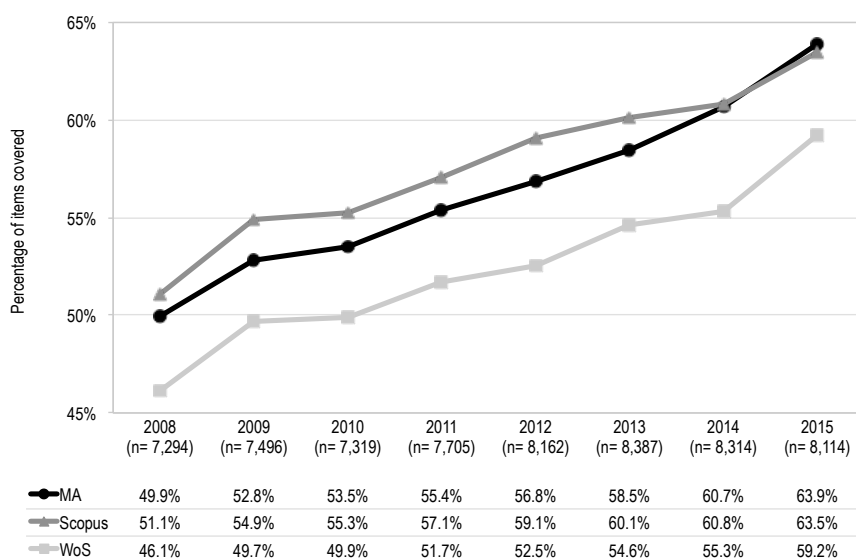

*Research field*

Field information from MA cannot readily be used for classifying publications into disciplines (Hug et al., 2017). Hence, we used the *Revised Field of Science and Technology (FOS) Classification* of the Frascati Manual (OECD, 2007) and assigned each of the 139 institutes at UZH to one FOS field. For each publication, the FOS field from the affiliated institute(s) as registered in ZORA was adopted. If a publication was assigned to two or more fields, the publication was counted and analyzed in each field. For the 62,791 publications in ZORA$^{2008-2015}$, 65,445 assignments were made, which resulted in an average of 1.04 FOS fields per publication. We analyzed the coverage of the six major FOS fields and examined the social sciences, humanities, and natural sciences more closely. We focused on these three fields, as publication coverage is an issue in many subfields of the social sciences and humanities (Gumpenberger, Sorz, Wieland, & Gorraiz, 2016; Mongeon & Paul-Hus, 2016) as well as in two subfields of the natural sciences, mathematics and computer/information sciences (Bosman, van Mourik, Rasch, Sieverts, & Verhoeff, 2006; Larsen & von Ins, 2010).

With respect to the coverage of major FOS fields, the three databases perform almost equally (see Fig. 4). The databases cover ZORA$^{2008-2015}$ publications in the social sciences and humanities poorly but perform very well in the other major FOS fields. The largest differences between MA and the benchmark databases are in agricultural sciences (MA: 64.7%; Scopus: 71.6%) and social sciences (MA: 33.7%; WoS: 26.5%).

**Fig. 4**   Coverage of ZORA$^{2008-2015}$ by major FOS fields

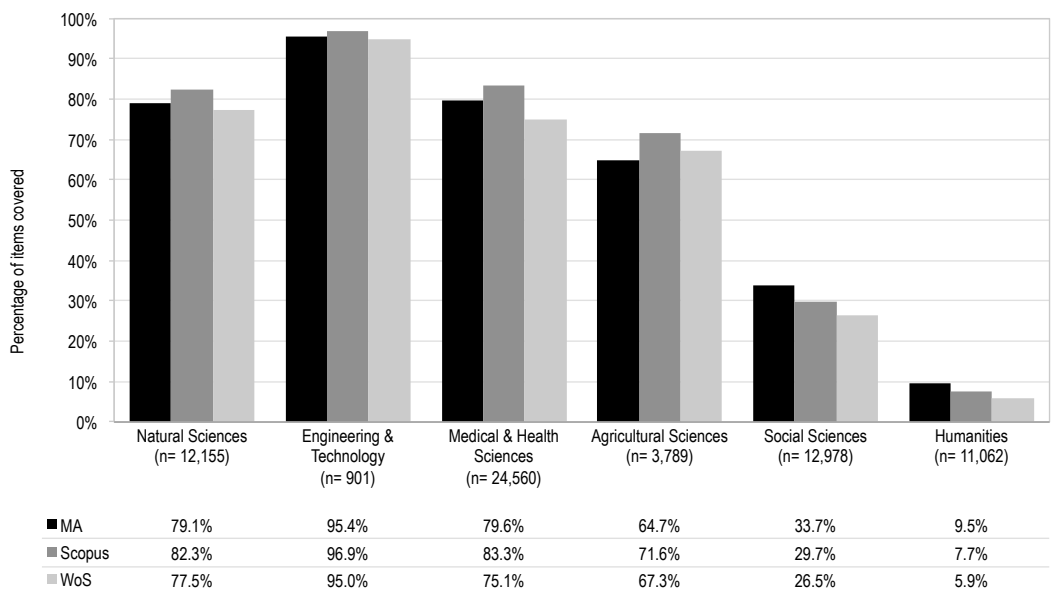

With regard to subfields of the natural sciences, MA covers at least two-thirds of the ZORA$^{2008\text{-}2015}$ publications in each subfield (see Fig. 5). The three databases cover biological sciences and mathematics almost equally. Given that previous studies have identified coverage issues in mathematics (Bosman et al., 2006; Larsen & von Ins, 2010), the coverage of mathematics publications in our sample was unexpectedly high in all three databases. We consulted several mathematicians at UZH, who suggested that this relatively high coverage might be due to the tendency of UZH mathematicians to mainly publish in highly ranked English journals. Due to a better coverage of conference items, MA and Scopus clearly outperform WoS in computer/information sciences as well as in other natural sciences. MA also outperforms WoS in earth and environmental sciences due to better coverage of conference items and book sections. Scopus and WoS cover chemical and physical sciences almost perfectly while MA shows a significantly lower coverage in these two subfields.

We analyzed the ZORA$^{2008\text{-}2015}$ publications in chemical and physical sciences that are not covered by MA (199 and 463 items, respectively) and found that 140 publications in chemistry and 374 publications in physics have titles with complex technical terminology that includes non-alphanumeric characters, symbols, punctuation marks, Greek letters, or mathematical expressions in LaTeX format. We manually searched some of these publications via the MA search engine and were able to find them. However, this required considerable effort since the use of technical terms in titles is inconsistent within and between original publications, ZORA, and MA. For example, the first part of the chemical ligand name "$\eta^6$-arene" comes in many variations, such as $\eta^6$, $\eta$ 6, $\eta$ 6, $\eta$ (6), or eta6. These and more complex cases prevented the Perl script from finding publications in MA and caused the significantly lower coverage of MA in chemistry and physics. If these issues related to title search could be resolved, the Perl script could also retrieve ZORA$^{2008\text{-}2015}$ publications with complex technical terms in the title (140 items in chemistry and 374 items in physics). This would increase the coverage of MA from 83.7% to 95.2% (+11.5%) in chemistry and from 67.5% to 93.8% (+26.3%) in physics. Hence, MA would cover these two subfields to a similar extent as Scopus and WoS. The total coverage of ZORA$^{2008\text{-}2015}$ would increase from 56.6% to 57.4% (+0.8%). However, the issues described above cannot easily be resolved. For this reason, specialized chemical databases such as SciFinder or Reaxys have been built, and specific search procedures such as structure and reaction search are needed in the field of chemistry (Currano & Roth, 2014). For bibliometric purposes, however, it would suffice if Microsoft implemented DOI queries in MA.

**Fig. 5** Coverage of ZORA[2008-2015] by FOS subfields of the natural sciences

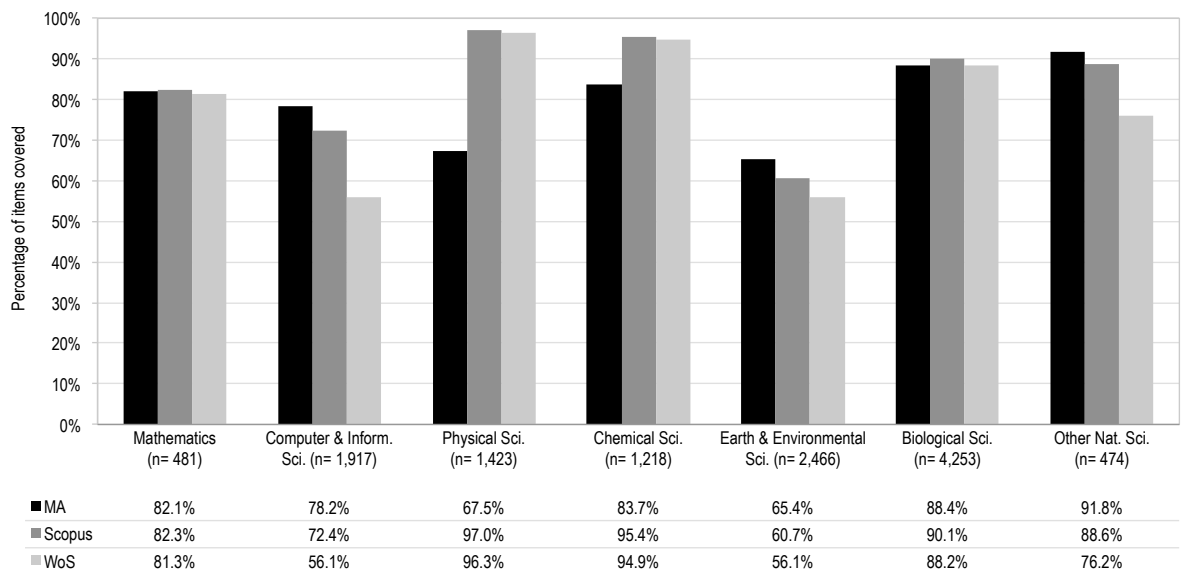

| | Mathematics (n= 481) | Computer & Inform. Sci. (n= 1,917) | Physical Sci. (n= 1,423) | Chemical Sci. (n= 1,218) | Earth & Environmental Sci. (n= 2,466) | Biological Sci. (n= 4,253) | Other Nat. Sci. (n= 474) |
|---|---|---|---|---|---|---|---|
| MA | 82.1% | 78.2% | 67.5% | 83.7% | 65.4% | 88.4% | 91.8% |
| Scopus | 82.3% | 72.4% | 97.0% | 95.4% | 60.7% | 90.1% | 88.6% |
| WoS | 81.3% | 56.1% | 96.3% | 94.9% | 56.1% | 88.2% | 76.2% |

In contrast to the subfields of the natural sciences, humanities subfields are covered very poorly by the three databases (see Fig. 6). This is due to different publication patterns in the natural sciences and the humanities. While in ZORA[2008-2015] the natural sciences feature journal articles, conferences items, and publications in English, which are very well covered by the three databases, the humanities feature book-related items and non-English publications, which are hardly covered by MA, Scopus, and WoS (see Fig. 1 and Table 5). MA does not substantially outperform the benchmark databases in any of the humanities subfields.

**Fig. 6** Coverage of ZORA[2008-2015] by FOS subfields of the humanities

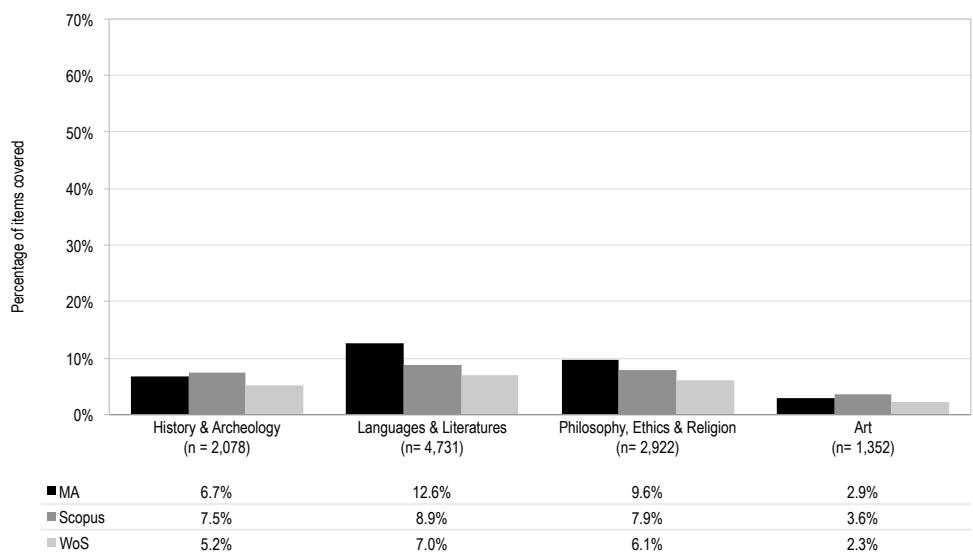

| | History & Archeology (n= 2,078) | Languages & Literatures (n= 4,731) | Philosophy, Ethics & Religion (n= 2,922) | Art (n= 1,352) |
|---|---|---|---|---|
| MA | 6.7% | 12.6% | 9.6% | 2.9% |
| Scopus | 7.5% | 8.9% | 7.9% | 3.6% |
| WoS | 5.2% | 7.0% | 6.1% | 2.3% |

The three databases cover ZORA[2008-2015] publications in subfields of the social sciences (see Fig. 7) either to a high degree (psychology, economics/business), a medium degree (sociology, political sciences), or low degree (educational sciences, law, media/communications, other social sciences). Overall, MA shows similar strengths and weaknesses as Scopus and WoS in covering social sciences subfields. However, MA is less biased, as it outperforms the benchmark databases in many subfields due to a broader coverage of journal articles and book-related items in the respective fields. In particular, MA clearly outperforms WoS in political sciences and in economics/business. When taking working papers into account as well, which are characteristic for economics/business and which MA covers to a high degree (see Fig. 2), MA outperforms Scopus and WoS by a large margin in economics/business. MA also outperforms one or both of the benchmark databases in educational sciences, law, media/communications, and in other social sciences, although to a lesser extent. The three databases cover psychology and sociology almost equally.

**Fig. 7** Coverage of ZORA[2008-2015] by FOS subfields of the social sciences

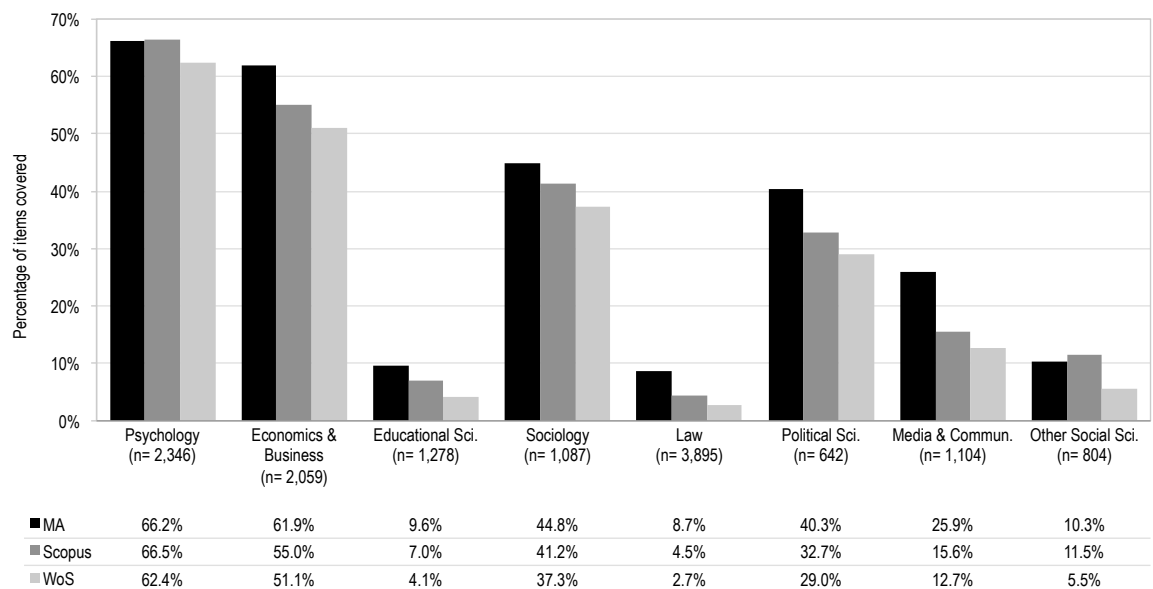

| | Psychology (n= 2,346) | Economics & Business (n= 2,059) | Educational Sci. (n= 1,278) | Sociology (n= 1,087) | Law (n= 3,895) | Political Sci. (n= 642) | Media & Commun. (n= 1,104) | Other Social Sci. (n= 804) |
|---|---|---|---|---|---|---|---|---|
| MA | 66.2% | 61.9% | 9.6% | 44.8% | 8.7% | 40.3% | 25.9% | 10.3% |
| Scopus | 66.5% | 55.0% | 7.0% | 41.2% | 4.5% | 32.7% | 15.6% | 11.5% |
| WoS | 62.4% | 51.1% | 4.1% | 37.3% | 2.7% | 29.0% | 12.7% | 5.5% |

**Citation counts**

MA offers two citation measures: estimated citation counts (attribute ECC in the AK API) and citation counts (attribute CC), which are calculated based on linked references in the MAG (A. Wade, personal communication, September 1, 2016). We used citation counts for two reasons. First, calculating citation counts based on linked references is the standard method in bibliometrics and is employed in Scopus and WoS. Second, Microsoft has not fully disclosed the method for calculating estimated citation counts (for a sketch of the method, see Harzing & Alakangas, 2017a).

*Citations per publication, uncitedness*

The ZORA$^{2008-2015}$ publications collected 745,758 citations in Scopus, 669,084 citations in WoS, and 652,081 citations in MA. The three databases show almost identical citations per publication (CPP) in five of the six FOS fields (see Fig. 8). The humanities are the only FOS field with very low CPP values. In the natural sciences, CPP values are about a quarter lower in MA than in Scopus and WoS. Within the natural sciences, CPP values of MA, Scopus, and WoS are similar in four subfields but differ considerably in physical sciences (14.4 vs. 41.5 and 43.0 CPP, respectively), chemical sciences (14.8 vs. 20.6 and 20.1 CPP, respectively), and computer/information sciences (11.2 vs. 23.2 and 29.4 CPP, respectively). A detailed analysis of publications in these three subfields revealed that most of the publications have lower or considerably lower citations in MA than in Scopus and WoS. Only a few publications have more (but none have considerably more) citations in MA. The share of uncited ZORA$^{2008-2015}$ publications in WoS (12.2%; 4,016 of 33,000 items) is slightly lower than in MA (16.1%; 5,720 of 35,557 items) and Scopus (15.2%; 5,536 of 36,351 items).

**Fig. 8** Citations per publication of ZORA$^{2008-2015}$ items by FOS field

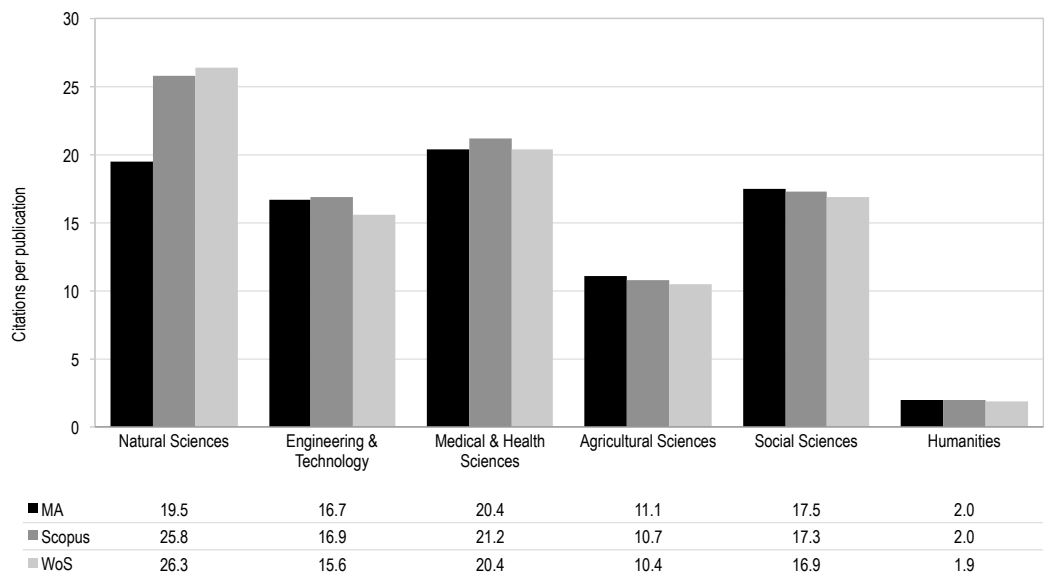

| | Natural Sciences | Engineering & Technology | Medical & Health Sciences | Agricultural Sciences | Social Sciences | Humanities |
|---|---|---|---|---|---|---|
| MA | 19.5 | 16.7 | 20.4 | 11.1 | 17.5 | 2.0 |
| Scopus | 25.8 | 16.9 | 21.2 | 10.7 | 17.3 | 2.0 |
| WoS | 26.3 | 15.6 | 20.4 | 10.4 | 16.9 | 1.9 |

*Correlation of citations*

Citations of ZORA$^{2008-2015}$ publications were collected in the three databases and correlated using the Pearson correlation coefficient, Spearman's rho (mean rank for ties), and Kendall's Tau-b. Pearson's coefficients yield a very high correlation between Scopus and WoS ($r=0.99$, $p<0.01$) and lower but still high correlations between MA and Scopus ($r=0.73$, $p<0.01$) and between MA and WoS ($r=0.73$, $p<0.01$). This indicates that citation counts are more closely related between Scopus and WoS than between MA and the benchmark databases, which is also reflected in the respective scatterplots (see Fig. 9).

**Fig. 9** Relationship between citations of publications from ZORA$^{2008-2015}$ in MA, Scopus, and WoS

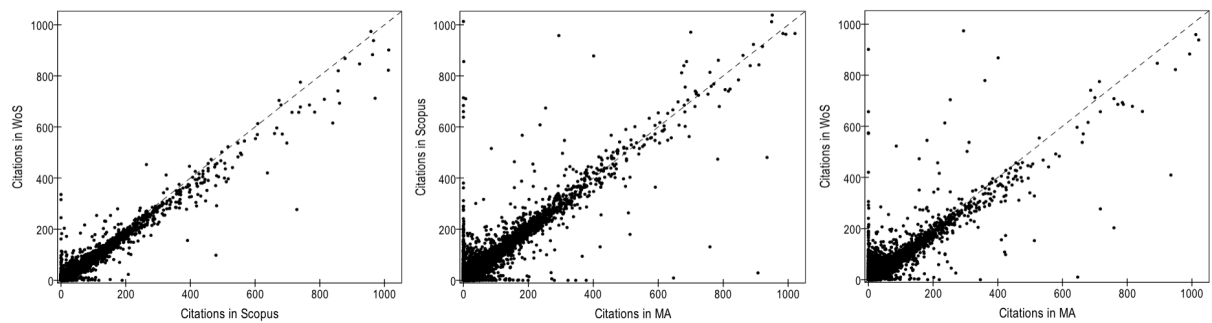

*Note:* Publications with more than 1,050 citations are not displayed. They represent less than 0.1% of the ZORA$^{2008-2015}$ publications covered by MA, Scopus, and WoS.

With respect to rank correlations, coefficients are very high between Scopus and WoS and slightly lower but still high between MA and the benchmark databases (see Table 6). These correlation patterns hold true for all FOS fields. The humanities are the only field with somewhat weaker correlations. Overall, this indicates that publications are ranked very similarly based on citations from MA, Scopus, and WoS.

Table 6   Rank correlations of citations of ZORA$^{2008-2015}$ publications by FOS fields

|  | MA / Scopus | | MA / WoS | | Scopus / WoS | |
| --- | --- | --- | --- | --- | --- | --- |
|  | $n$ | $r_s$ / $r_\tau$ | $n$ | $r_s$ / $r_\tau$ | $n$ | $r_s$ / $r_\tau$ |
| All fields | 32,164 | .90 | 29,960 | .89 | 31,880 | .96 |
|  |  | .80 |  | .79 |  | .89 |
| Natural Sciences | 8,849 | .86 | 8,356 | .85 | 9,148 | .96 |
|  |  | .74 |  | .72 |  | .89 |
| Engineering & Technology | 842 | .93 | 825 | .93 | 848 | .97 |
|  |  | .83 |  | .82 |  | .91 |
| Medical & Health Sciences | 18,678 | .93 | 17,377 | .93 | 18,194 | .97 |
|  |  | .84 |  | .84 |  | .90 |
| Agricultural Sciences | 2,310 | .93 | 2,246 | .93 | 2,503 | .96 |
|  |  | .83 |  | .83 |  | .90 |
| Social Sciences | 3,357 | .83 | 3,073 | .85 | 3,167 | .96 |
|  |  | .73 |  | .74 |  | .89 |
| Humanities | 400 | .73 | 308 | .74 | 347 | .89 |
|  |  | .65 |  | .64 |  | .88 |

*Note:* $n$ = number of items covered in two databases at the same time. $r_s$ = Spearman's rho (mean rank for ties). $r_\tau$ = Kendall's Tau-b. All correlations $p < 0.01$.

## Discussion

In this study, the coverage of MA was assessed based on the repository of UZH (ZORA) and compared with two benchmark databases (Scopus and WoS). In addition, citation counts were analyzed and issues related to data retrieval and quality were examined. As the DOI of a publication cannot be utilized to retrieve metadata from MA (Hug et al., 2017), the titles of publications were used. A Perl script was written that retrieves publications either based on the exact title (ti_ex mode) or based on title words (ti_wo mode) via the AK API. The script is freely available on GitHub. While the ti_ex mode and the ti_wo mode performed equally in terms of recall, ti_ex had a higher precision than ti_wo. Since each mode retrieved some items from MA that the other did not, we suggest using both modes to maximize the number of retrieved items. Almost all retrieved items that match publications from ZORA rank in the top three of the AK API result set (ti_ex: 99.4% of the matched publications; ti_wo: 98.4%). These results indicate that both modes translate title information into meaningful MA requests and that MA delivers very precise results. However, this does not hold true for publications from chemistry and physics since complex technical terms (e.g., non-alphanumeric characters, symbols, mathematical expressions in LaTeX format) prevented the Perl script from identifying some publications in MA. We assume that, due to this issue, the present study underestimates the coverage of MA by 11.5% in chemistry, by 26.3% in physics, and by 0.8% in the repository as a whole. To avoid this problem and facilitate the retrieval of MA metadata in general, DOI queries should be permitted in the AK API. According to the development team of MA, Microsoft is considering implementing DOI queries in the future (A. Chen, personal communication, March 31, 2017).

We found that 89.5% of the items retrieved from MA have correct publication years, 7.0% differ by ±1 year, and 3.5% feature larger differences. These results corroborate the findings of Herrmannova and Knoth (2016a), who reported almost identical numbers (i.e. 88%, 8%, and 4%, respectively). Microsoft (2017c) offers an explanation for differences in publication years, particularly for publications "made older" by MA: "Many publications are posted online before appearing in conferences or journals. As a result, MA tends to use the 'first seen' date as publication date." We could not find further explanations for differences in publication years (e.g., differences related to specific publishers, document types, or publication years) other than parsing errors revealed by spot checks (e.g., 1866 instead of 1966). With respect to authorship, our analysis showed that 95.1% of the retrieved journal articles list the correct number of authors, 1.7% differ by ±1 author, and 3.2% exhibit larger differences. According to our analysis, Microsoft limits the maximum number of authors per publication to 50 in the AK API. This would explain why there are occasionally fewer authors in MA. The availability of DOIs for different research fields in MA is comparable to those in Scopus and WoS reported by Gorraiz, Melero-Fuentes, Gumpenberger, & Valderrama-Zurian (2016). However, 12.6% of the publications that have a DOI in ZORA do not have one in MA.

Overall, MA and Scopus cover ZORA to a similar extent. Of the three databases, MA covers the most publications from ZORA exclusively. MA and Scopus both outperform WoS in terms of overall as well as unique coverage. While Scopus has a slightly better coverage of journal articles than MA and WoS, MA surpasses Scopus and WoS with respect to conference items and book-related document types (i.e., monographs, edited volumes, book sections). In addition, MA includes further document types (e.g., dissertations, habilitations, working papers, research reports), which are not indexed by Scopus and WoS (Clarivate, 2017; Elsevier, 2017). MA covers working papers, journal articles, and conference items from ZORA to a substantial extent. None of the newspaper articles from ZORA could be found in MA, which indicates that MA manages to separate scholarly from non-scholarly content. Performing this discrimination successfully proves to be difficult for academic search engines such as Google Scholar (Orduna-Malea, Ayllón, Martín-Martín, & López-Cózar, 2015).

For a detailed comparison of MA with the benchmark databases, the analysis was restricted to publications from 2008 to 2015 and to main document types used in scholarly communication and evaluative bibliometrics (i.e., journal articles, book-related items, conference items). This restricted publication set is referred to as $ZORA^{2008-2015}$. Regarding publication language and access status of full texts, the three databases perform similarly. They feature a high coverage of publications in English and a very low coverage of publications in languages other than English (mostly German, French, and Italian in ZORA). The three databases cover publications that are not publicly accessible to a high degree and open-access publications to a somewhat lower degree. Hence, MA exhibits similar language and open-access biases as previous studies have reported for Scopus and WoS (Mas-Bleda & Thelwall, 2016; Moed, Bar-Ilan, & Halevi, 2016).

With regard to research fields, our results suggest that the coverage of MA is favorable for evaluative bibliometrics in the natural sciences (including mathematics and computer/information sciences), engineering/technology, medical/health sciences, and in two subfields of the social sciences (psychology and economics/business). In these fields, MA covers two-thirds or more of the ZORA$^{2008\text{-}2015}$ publications. In further fields, MA covers ZORA$^{2008\text{-}2015}$ publications less favorably. In the social sciences, MA outperforms the benchmark databases in most subfields; however, it does only reach low (e.g., educational sciences, law) or medium (e.g., sociology, political sciences) coverage. In the humanities, the three databases perform poorly in all subfields, and MA does not outperform Scopus and WoS. Thus, except for in psychology and economics/business, MA shows similar biases with regard to the social sciences and humanities as previous studies have identified in Scopus and WoS (Gumpenberger et al., 2016; Mongeon & Paul-Hus, 2016).

We found that ZORA$^{2008\text{-}2015}$ publications are ranked very similarly based on citations from MA, Scopus, and WoS. In particular, the rank correlations between Scopus and WoS were *very* high, while the correlations between MA and the benchmark databases were high. This suggests that citation analyses with MA, Scopus and WoS should yield relatively similar results. Furthermore, we found that the three databases show almost identical citations per publication (CPP) with respect to a research field. However, this does not hold true for the natural sciences, where MA shows lower CCP values than the benchmark databases in the physical, chemical, and computer/information sciences. In these fields, most of the publications have lower or considerably lower citations in MA than in Scopus and WoS. Future studies need to assess if the observations in these three fields are robust or only apply to our sample.

Our findings suggest that, with the exceptions discussed above, MA performs similarly to Scopus in terms of coverage and citations. While this speaks for the quality of MA, it is at the same time somewhat deflating, as one expects superior performance from a database that gets most of its data from indexed web pages (Sinha et al., 2015). However, since MA is only in its second year and still developing, such expectations are perhaps misplaced. Due to the rapidly growing nature of MA, we expect future studies to find a much broader coverage of publications and considerably higher citation counts.

**Conclusion**

With its rapid and ongoing development, MA is on the verge of becoming a bibliometric superpower. MA offers rich and structured metadata and an API of high functionality (Hug et al., 2017). The present study and the studies of Harzing and Alakangas (2017a, 2017b) provide initial evidence for the excellent performance of MA in terms of coverage and citations. The features of MA are not only advantageous for bibliometric purposes but also for literature search (Hug & Brändle, 2017) and for library applications such as enriching institutional repositories or assessing library collection and acquisitions by using reference information.

Many academic search engines are rather opaque about the sources they cover (Ortega, 2014) and, according to Moed et al. (2016), this is particularly true for Google Scholar. Unfortunately, this also applies to MA. Despite the open approach taken by its development team, the only known sources of MA are metadata feeds from publishers and web pages indexed by Bing (Sinha et al., 2015). Hence, future studies need to address the source coverage of MA. Moreover, it would be helpful if Microsoft were to elaborate on its coverage policy as well as on how MA data are curated and cleaned. Our findings and conclusions depend on how open, transparent, and accessible MA data will remain in the future as well as on the quality of MA metadata, which have not yet been investigated in detail. Until MA metadata have been (positively) evaluated, we recommend using MA for evaluative bibliometrics in conjunction with established databases (e.g., Scopus, WoS, disciplinary database).

The present study is limited to the publications of a Swiss university in the period of 2008 to 2015. Further studies are needed to examine the coverage of different and broader publication sets (including publications in non-Western languages, as suggested by Fagan, 2017).

## Acknowledgments

The authors thank the development team of Microsoft Academic for their support, the ZORA editorial team for their advice, Robin Haunschild for comments, Mirjam Aeschbach for proofreading, and the reviewers for their remarks.